\documentclass[aps,pre,twocolumn,showpacs,floatfix]{revtex4-1}

\usepackage[pdftex]{graphicx}
\usepackage{amssymb,amsfonts,amsmath}
\usepackage{color}
\usepackage[usenames,dvipsnames]{xcolor}
\usepackage{sidecap}

\usepackage{mhchem}
\usepackage{times,mathptmx}
\usepackage{balance} 

\usepackage{lastpage}
\usepackage[format=plain,justification=raggedright,singlelinecheck=false,font=small,labelfont=bf,labelsep=space]{caption} 
\usepackage{fancyhdr}
\usepackage{hyperref}
\usepackage{latexsym}
\usepackage{textcomp}
\usepackage{bm}

\usepackage[normalem]{ulem}

\newcommand*{\citen}[1]{%
  \begingroup
    \romannumeral-`\x 
    \setcitestyle{numbers}%
    \cite{#1}%
  \endgroup   
}

\begin{document}

\title{Contrasting the dynamics of elastic and non-elastic deformations across an experimental colloidal Martensitic transition}
\author{Saswati Ganguly$^{\ast}$\textit{$^{a}$}, Priti S. Mohanty\textit{$^{b,e}$}, Peter Schurtenberger\textit{$^{e}$}, Surajit Sengupta\textit{$^{c}$}, and Anand Yethiraj\textit{$^{d}$}}
\affiliation{$^{a}$~Institut f\"ur Theoretische Physik II: Weiche Materie, Heinrich 
Heine-Universit\"at D\"usseldorf, Universit\"atsstra{\ss}e 1, 40225 D\"usseldorf, Germany. Email: saswati@thphy.uni-duesseldorf.de\\
$^{b}$~School of Applied Sciences, KIIT University, Bhubaneswar 751024, India. Email: pritimohanty02@gmail.com\\
$^{c}$~TIFR Centre for Interdisciplinary Sciences, 21, Brundavan Colony, Narsingi, Hyderabad 500075, India. Email: surajit@tifrh.res.in\\
$^{d}$~Department of Physics and Physical Oceanography, Memorial University, St. John's, Newfoundland Labrador, A1B 3X7, Canada. Email:  ayethiraj@mun.ca\\
$^{e}$~Division of Physical Chemistry, Lund University, SE-221 00 Lund, Sweden. Email:  Peter.Schurtenberger@fkem1.lu.se}

\begin{abstract}
We present a framework to segregate the roles of elastic and non-elastic deformations in the examination of real-space experiments of solid-solid Martensitic transitions. The Martensitic transformation of a body-centred-tetragonal~(BCT) to a body-centred-orthorhombic~(BCO) crystal structure has been studied in a model system of micron-scale ionic microgel~\cite{Priti} colloids. Non-affine fluctuations, {\it i.e.,}  displacement fluctuations that do not arise from purely elastic~(affine) deformations, are detected in particle configurations acquired from the experiment. Tracking these fluctuations serves as a highly sensitive tool in signaling the onset of the Martensitic transition and precisely locating particle rearrangements occurring at length scales of a few particle diameters. Particle rearrangements associated with non-affine displacement modes become increasingly favorable during the transformation process. The nature of the displacement fluctuation modes that govern the transformation are shown to be different from those predominant in an equilibrium crystal. 
We show that BCO crystallites formed through shear may, remarkably, co-exist with those resulting from local rearrangements {\em within the same sample}. 
\end{abstract}

\maketitle
\section{Introduction}
\label{intro}
Martensitic transformations are characterized by the coordinated motion of particles over long particle length scales leading to a diffusion-less phase transition between solids. They occur in a variety of materials like metals, alloys, ceramics and proteins, and have practical consequences in a range of phenomena,  from work hardening of metals~\cite{porter_phase_2009,bhadeshia_steels:_2011} and ceramics~\cite{pelleg_strength_2014}  to the ``shape memory effect'' in alloys~\cite{otsuka_shape_1998}.

Martensitic transitions are governed by the strain energies associated with the homogeneous lattice-distortive strains and the energies arising due to formation of the parent-product interfaces. The homogeneous lattice deformation or the Bain strain~\cite{Bain}, makes an elastic order parameter a natural choice for understanding such transformations. {\color{black} However, the constraint of being forced to grow a product nucleus within the parent matrix necessitates  ``lattice invariant" deformations such as slips or twinning in order to minimize total energy. The exact nature of these deformations depend on the structure and the elastic properties of both the parent and product crystals. Transformations associated with practically no plastic deformations, like the ones observed in shape memory alloys, are reversible. Quenching of steel, on the other hand, results in an irreversible transformation often involving plastic deformation.} Analyzing Martensitic transformations in terms of an elastic order parameter~\cite{Barsch,Lookman,Lookman_Shenoy1,Gooding}, imposing local elastic compatibility conditions, has the restriction of smooth displacement fields which do not allow dislocations or slips. 

The factors affecting a structural transition and the wide array of interesting micro-structural changes associated with it cannot be fully understood within the scope of elasticity theories. Studies done for the transition from square to rhombic {\color{black}crystal} structures indicate that the dynamics of non-elastic displacements, associated with plastic events, have a significant influence in determining the microstructure of the product phase~\cite{Joyee1,Joyee2,Arya,Rao_Sengupta,peng_two-step_2014}. External factors such as rate and depth of quench, leading to the structural transitions, are known to influence the resulting microstructure. Such external variables are correlated to the non-elastic displacement variables. 

It is difficult to study kinetic pathways of Martensitic transformation in atomic solids because of length and  timescale limitations as well as lack of access to particle trajectories. Colloids are attractive model systems because of the detailed real-space information readily  obtainable at the single-particle (``atomic'') level. This has been used to obtain new insights into novel crystal structures~\cite{pusey_phase_1986,monovoukas_gast_1989,halsey_electrorheological_1992,gast_simple_1998,dassanayake_structure_2000,yethiraj_colloidal_2003}, hexatic phases \cite{zahn_dynamic_2000,marcus_observations_1996}, crystal-to-crystal and crystal-to-glass transitions~\cite{weiss_martensitic_1995, yethiraj_prl_2004, mohanty_deformable_2012} as well as multiple glassy states in a single colloidal system~\cite{Pham_2002}. The timescales of colloidal motions can, in principle, also enable studies of phase transformation kinetics in real time. There have been pioneering studies of colloidal kinetics but they have been done either in 2 dimensions~\cite{grier_murray_1994}, or for slow dynamics in glassy systems~\cite{gasser_real-space_2001,Kegel_2000,Weeks_2000}, or in reciprocal space methods~\cite{palberg_crystallization_1999}. Recent advances in confocal imaging methods have resulted in a flurry of experimental work studying phase transformation kinetics of colloidal solids via confocal laser scanning microscopy(CLSM)~\cite{Priti, yang, peng, jenkins}. 
{\color{black}While there has been much theoretical work dedicated to studying solid-solid phase transformations ~\cite{Barsch,Lookman,Lookman_Shenoy1,Gooding,Joyee1,Joyee2,Arya,Rao_Sengupta,PhysRevLett.115.185701}, there has been no coherent comparison of the colloids experiments with theory.}

In this work, {\color{black}we utilize a theoretical framework to segregate elastic and non-elastic displacements and }show that {\em both}  play essential roles in the formation of the product crystal for a colloidal Martensitic transition.
It was shown in~[\citen{Priti}], by explicitly scanning in 3 dimensions, that turning on an electric field at high enough particle volume fractions resulted in a body-centred tetragonal (BCT) crystal with $c$-axis along the field ($z$) direction, while turning off the field resulted in a Martensitic transformation to a metastable body-centred orthorhombic (BCO) crystal, also with $c$-axis along the $z$ direction.
The transformation from a BCT to a BCO crystal structure in ~[\citen{Priti}] is a 3-dimensional crystal phase transformation.
Nevertheless, because the $c$-axis of both parent and daughter crystals was {\color{black} parallel} to the viewing direction, the dynamics could be followed in a quasi-2D study.

\begin{figure}[h!]
\begin{center}
\includegraphics[width=0.49\textwidth]{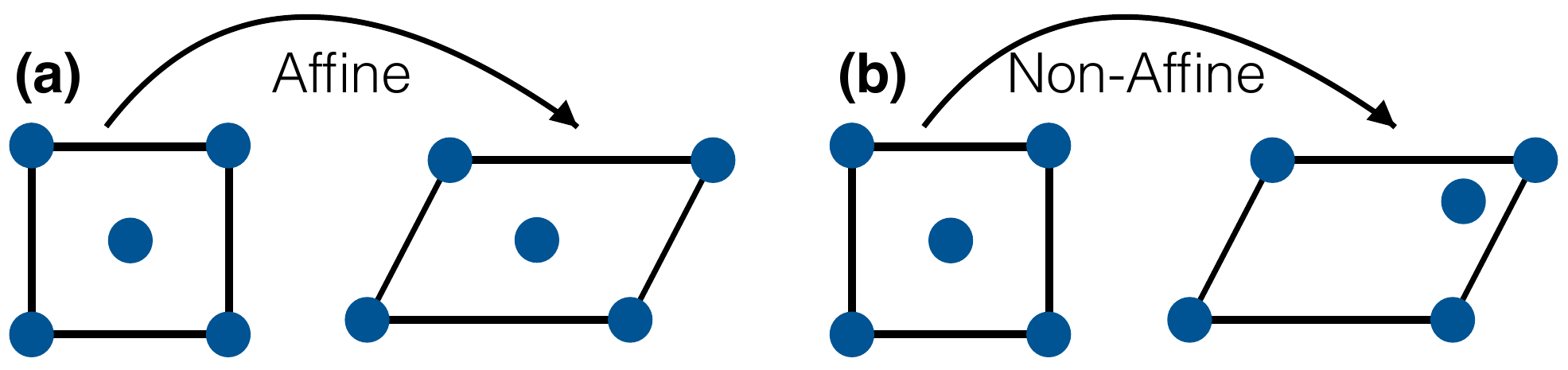}
\caption{\label{ANA}{\bf a.} Shows an {\it affine} shear transformation. A single deformation tensor can represent the displacements of all the particles. {\bf b.} Shows a {\it non-affine} displacement transformation. The displacements of the particle at the center of the box cannot be represented by the shear matrix alone that gives the displacements of the particles at the vertices of the box.}
\end{center}
\end{figure}

We study the displacement fluctuations along the BCT to BCO colloidal Martensitic transformation path using the language of affine and non-affine displacements~\cite{SG1,SG2}. Affine displacements (Fig~\ref{ANA}{\bf a}) of particles are the elastic part of the response of a solid to an external or internal stress. Non-affine displacements(Fig~\ref{ANA}{\bf b}), on the other hand, cannot be described in terms of { elastic strains alone and may represent particle rearrangements}. 

We project the displacements onto mutually orthogonal subspaces thus separating the purely affine strains from local {\it non-affineness}. This allows us to quantify and study the distinctive roles played by these displacement parameters in the colloidal BCT to BCO transformation. The study also makes it possible to illustrate a connection between the local elastic environment in the parent crystal and the distinctive displacement modes leading to crystallites in the product phase.

The paper is organized in the following sections. In Section~\ref{sec1A} we briefly describe the experiments performed~\cite{Priti}. { Section~\ref{sec1B} concerns the method of analysis of the data. Here} we describe the projection formalism to project particle displacements onto the affine and non-affine sub-spaces thus evaluating the total non-affinity, $\chi$ and the affine deformation, $\mathsf{D}$ associated with a particle within a predetermined coarse-graining volume. We present the displacement modes corresponding to the non-affine subspace. We analyse the experimentally obtained particle coordinates with the aforesaid framework and discuss our results in Section~\ref{sec2}. Section~\ref{sec2A} tests the validity of $\chi$ as an effective diagnostic tool in tracking the transition. In Section~\ref{sec2B}, we calculate the probability density functions of the susceptibilities of the non-affine displacements and the evolution of the gap separating the softest defect precursor modes with the other non-affine modes  at various stages of the transformation. In Section~\ref{sec2C} we discuss how different particle rearrangements during the Martensitic transformation can lead to identical crystal structures in the product phase. Finally we conclude in Section~\ref{sec3} with suggestions of future directions.

\section{Experiment and Analysis}
\subsection{Experiment}
\label{sec1A} For completeness, we briefly describe the experiments carried out in~[\citen{Priti}]. Confocal microscopy experiments of 2-dimensional ($x-y$) slices in the bulk of a 3D colloidal crystal were carried out in order to follow 2-dimensional trajectories of particles in real time. An a.c. electric field was applied perpendicular to the image plane (i.e, along $z$), resulting in the formation of particle chains along the field direction. Due to the quasi-2D nature and the relatively slow kinetics from BCT to BCO, this experimental work enabled the study of the detailed kinetics of a truly 3D crystal phase transformation with 2D trajectories.

\subsection{Affine and Non-affine displacements}
\label{sec1B}
The formalism described in this section is developed and discussed in detail in references~[\citen{SG1}] and~[\citen{SG2}]. Here we briefly discuss the ideas most relevant in the context of this work.

Displacements are identified as {\it affine} (Fig~\ref{ANA}{\bf a.}) or {\it non-affine} (Fig~\ref{ANA}{\bf b.}) with respect to a reference configuration of $N$ particles on which a particle
with index $i$ ($i=1,...,N$) is located at the position ${\bf R}_i$. Here the BCT structure at the beginning of the experiment ($time=0$) is considered as the reference. Since the experiments reported were of 2-dimensional slices of 3D crystals, the reference structure is a square lattice, and our analysis is 2D; however, the analysis can be carried out for 3D data as well. In the configurations in subsequent time steps, the displacement of particle $i$ from its position on the reference lattice
is given by ${\bf u}_i = {\bf r}_{i}-{\bf R}_{i}$, with ${\bf r}_i$ the
instantaneous position of the particle. Now within a neighborhood $\Omega$
around particle $i$, we define relative particle displacements $\bm{
\Delta}_{j} = {\bf u}_j-{\bf u}_i$ with particle $j\neq i \in\Omega$. The ``best fit'' \cite{falk} local affine deformation ${\mathsf D}$ is the
one that minimizes  $\sum_j [\bm{\Delta}_{j} - {\mathsf D}({\bf R}_{j}
- {\bf R}_{i})]^2$. The minimum value of this quantity,  $\chi({\bf R}_i) \ge 0$, is identified as the non-affinity parameter.
This is obtained by projecting~\cite{SG1} $\bm{\Delta}_i$ onto a non-affine
subspace defined by the projection operator ${\mathsf P}$ such that,
$\chi({\bf R}_i)= \bm{\Delta}^{\rm T}{\mathsf P}\bm{\Delta}$ and ${\mathsf D}={\mathsf Q}\bm{\Delta}$. $\Delta$ is the column vector constructed out of the  $\bm{\Delta}_i$. The projection operator ${\mathsf P}=I - R{\mathsf Q}$ with $\mathsf Q=(R^{T}R)^{-1}R^{T}$ and $R$ is the column vector constructed out of ${\bf R}_i$. The strain component of the affine part of the displacements is defined as $\varepsilon_s = (D_{xy} + D_{yx})/2$ where $D_{xy}, D_{yx}$ are the off-diagonal terms of the deformation matrix ${\mathsf D}$.

\begin{figure}[h!]
\begin{center}
\includegraphics[width=0.495\textwidth]{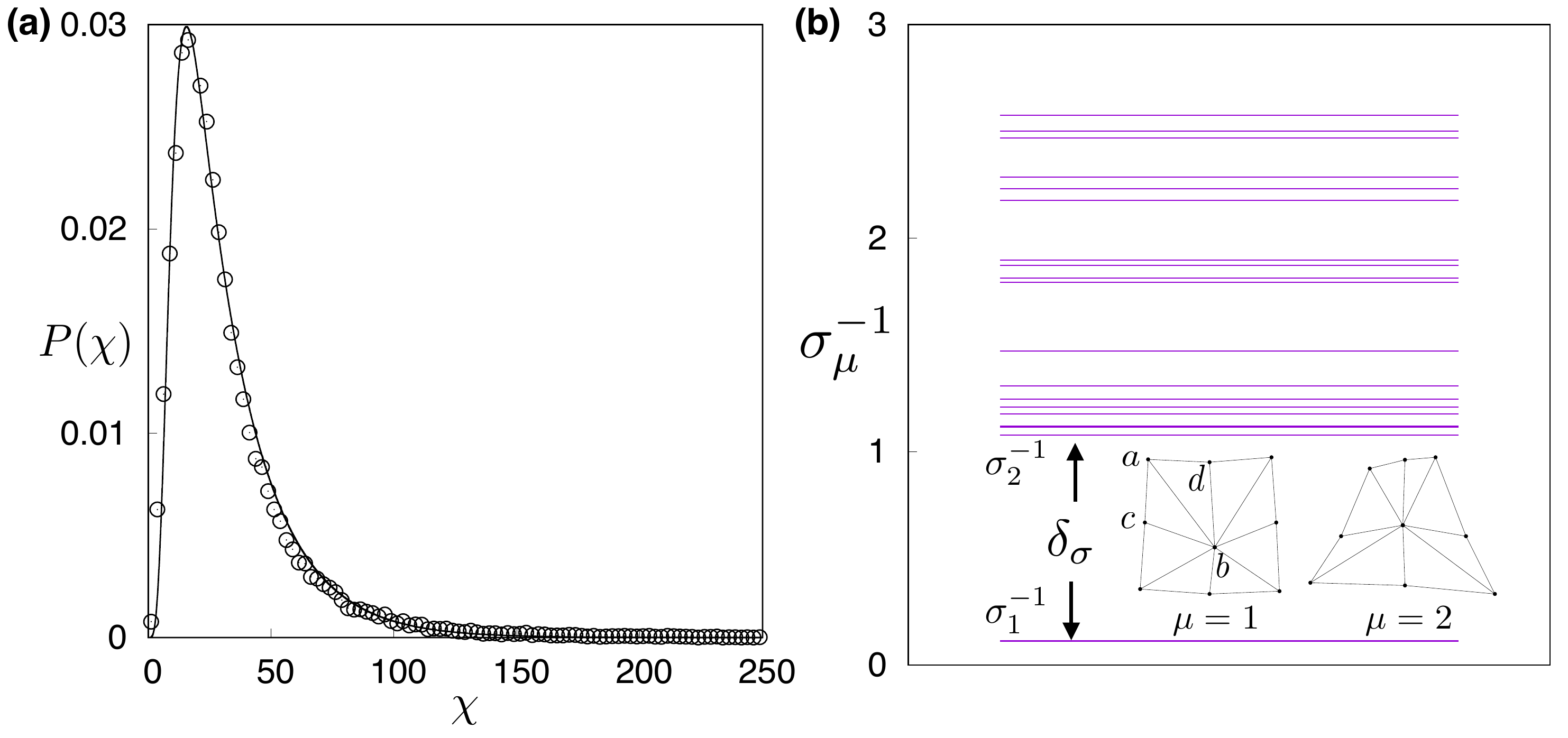}
\caption{\label{Ev_Harm}{\bf a.} Comparison of the probability distribution, $P(\chi)$ obtained from the 2D confocal images of the cross sections of the parent BCT crystal (black open circles) and the model harmonic square lattice network (black solid line) at $T=2.3$. {\bf b.} Plot of $\sigma^{-1}_{\mu}$ for the model harmonic square lattice structure. $\sigma^{-1}_{1} $ and $\sigma^{-1}_{2}$ are the two softest non-affine eigenmodes.  Note that the displacement eigenvector corresponding to $\sigma_{1}^{-1}$ has the vertices {\bf a-b} moving apart while the vertices {\bf c-d} coming closer, thus forming a dislocation precursor mode. $\delta_{\sigma}=\sigma_{2}^{-1}-\sigma_{1}^{-1}$, having dimensions of {\color{black} the square of inverse length}, estimates the gap between the two non-affine modes.}
\end{center}
\end{figure}

It has been shown in Refs.~[\citen{SG1,SG2}] that the non-affine displacement modes are the eigen-vectors corresponding to the non-zero eigenvalues of the {covariance matrix ${\mathsf C} =\langle \bm{\Delta}^{\rm T} \bm{\Delta}\rangle$ projected on the non-affine subspace. 
Thus  $\chi({\bf R}_i)$ is given by the sum of the eigenvalues ($\sigma_{\mu}$) of ${\mathsf P}{\mathsf C}{\mathsf P}$. The non-affine eigenvalues $\sigma_{\mu}$ can be interpreted as the susceptibilities associated with specific non-affine modes~\cite{SG2}. Thus, a smaller value of $\sigma_{\mu}^{-1}$ implies a softer mode. \color{black}{Indeed, within a harmonic theory, each $\sigma_{\mu}$ is simply the inverse curvature of the free energy in the direction of that specific eigenmode -- a direct consequence of equipartition.}  

 {\color{black} The two dimensional confocal images of the cross sections of the parent BCT crystal can be modeled as a square lattice network of particles connected by harmonic springs, {\color{black}$U=\frac{1}{2}K(r-r_{0})^{2}$}, at temperature $T=2.3$, spring constant $K=1.0$ and lattice parameter $l=1.0$ in reduced units implying that $T=\frac{T^{\prime}k_{B}}{\text{unit of energy}}$, $K=\frac{K^{\prime}} {\text{unit of energy}/(\text{unit of length})^{2}}$ and $l=\frac{l^{\prime}}{\text{unit of length}}$. $k_{B}$ is the Boltzmann constant and $T^{\prime}, K^{\prime}$ and $l^{\prime}$ are temperature, spring constant and length in conventional units. The length and the energy scales are set by the lattice parameter $l$ and $Kl^{2}$ respectively. This {\color{black}``$l$"} of the harmonic model is essentially equivalent to the diameter, $d=1 \ micron$ of the colloidal particles in our experiments. We chose to represent all the lengths in units of the particle diameter, $d$. $\chi$, defined in terms of displacement correlation matrix ${\mathsf C}$, naturally has the dimensions of length square and has been represented in units of $d^{2}$.   The probability distribution, $P(\chi)$ of the non-affine parameter obtained from the ideal harmonic network and the $P(\chi)$ calculated from the 2D cross section of the quiescent BCT crystal~(Fig.\ref{Ev_Harm}{\bf a}) shows reasonably good agreement.} To obtain a stable square lattice one needs nearest-neighbour as well as next-near-neighbour (diagonal) bonds. In the harmonic model we represent all the bonds by harmonic springs with the same spring constant $K$. The $\Omega$ chosen for calculating $\chi$ is slightly larger than the harmonic interaction volume. This corresponds to three nearest neighbor shells instead of two.
Since there are $13$ particles in the elementary volume $\Omega$, there are $12\times2 = 24$ displacement modes. Subtracting out the $4$ modes of affine deformation in two dimensions, one has $20$ non-affine modes and therefore $20$ non-trivial eigen-values of ${\mathsf P}{\mathsf C}{\mathsf P}$. As previously mentioned, the trace of the matrix ${\mathsf P}{\mathsf C}{\mathsf P}$ gives $\chi$ and its eigenvalues $\sigma_{\mu}$s have the same dimension as $\chi$. Therefore $\sigma_{\mu}^{-1}$ is represented in units of $d^{-2}$.
 Fig.~\ref{Ev_Harm}{\bf b.} shows the $\sigma^{-1}_{\mu}$ for the ideal square lattice.  In a two dimensional triangular lattice~\cite{SG2}
the eigen-vector associated with the most probable non-affine mode has been shown to be identical to thermally induced defect precursor modes. The situation is similar here. For example, in the $\mu = 1$ mode shown in Fig.~\ref{Ev_Harm}, the vertices a-b move apart while the c-d move closer to each other thus attempting to introduce a defect pair.
For ideal lattices at a finite temperature those especially soft modes are separated from other non-affine modes~\cite{SG2} by a relatively large gap $\delta_{\sigma}=\sigma_{2}^{-1}-\sigma_{1}^{1}$. The values of the $\sigma_{\mu}^{-1}$(Fig.~\ref{Ev_Harm}{\bf b.}),  represents the expectation values of the $\sigma_{\mu}^{-1}$ at the thermodynamic limit of an ideal crystal with no defects. In a finite sized real crystal at a given temperature, one would obtain a distribution of $\sigma_{\mu}^{-1}$ where the width of the distribution, centered around $\sigma_{\mu}^{-1}$, is determined by the temperature, system size, defect concentrations etc. The $\delta_{\sigma}$, given here, also provides an upper limit for the gap expected between the non-affine modes in a real system.} Finally, a large $\delta_{\sigma}$ implies that the non-affine component of all displacement fluctuations are dominated by the defect precursor mode while a small gap would mean that {\em all} non-affine modes with all possible non-strain distortions of $\Omega$ are equally important. 


\section{Results}
\label{sec2}
\subsection{$\chi$ { as a local parameter to track structural transformation kinetics}}
\label{sec2A}

{\color{black} We define a 2D local bond order parameter $\psi_{s}=|\frac{1}{N}\sum_{j=1}^{N}e^{i\theta j}|$ in order to quantify the local crystalline order of the transforming confocal images.}
\begin{figure}[h!]
\begin{center}
\includegraphics[width=0.49\textwidth]{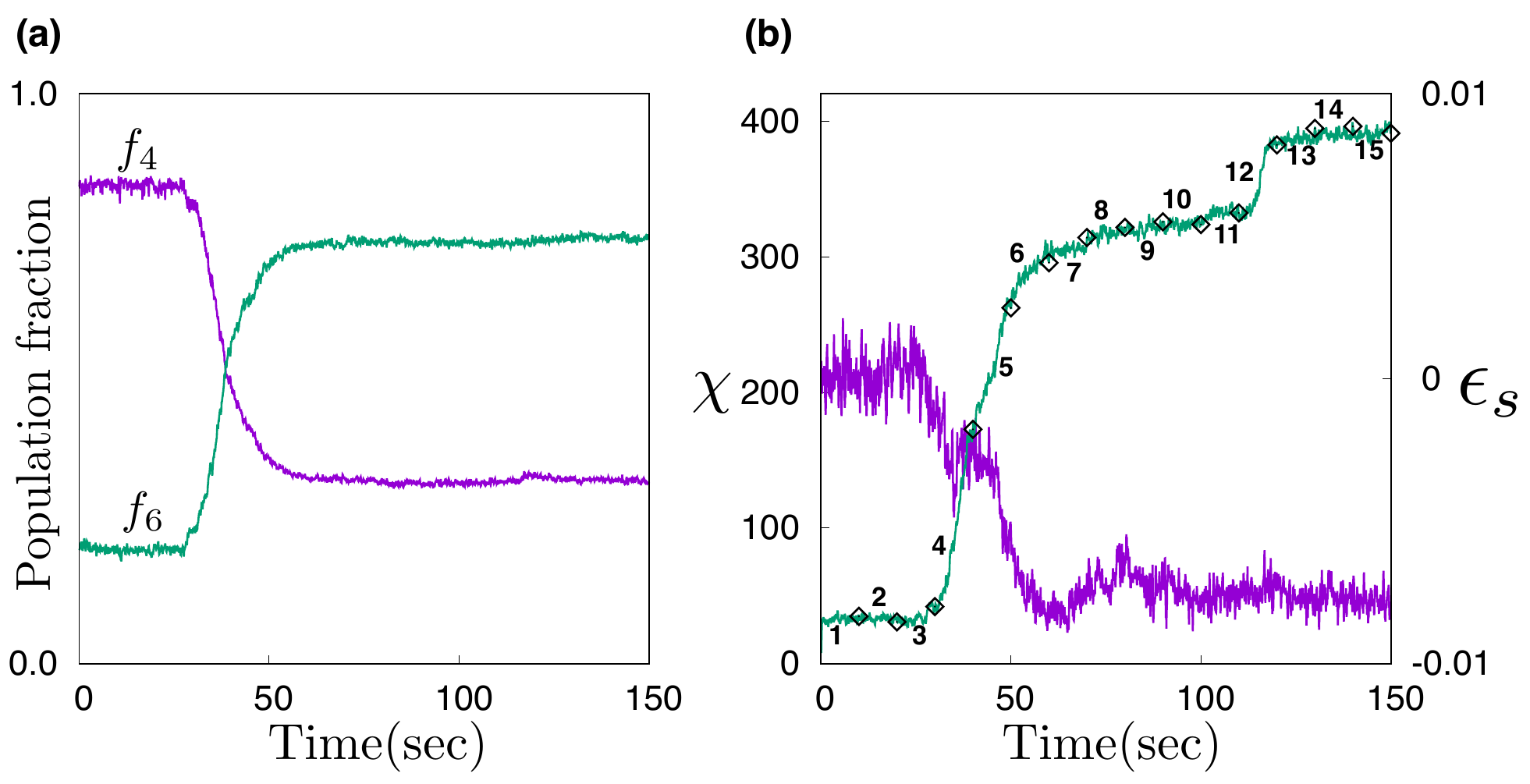}
\caption{\label{Op_time}{\bf a.} Plots of population fraction $f_{4}$ and $f_{6}$ as they evolve with time during the BCT to BCO transition. $f_{4}$ represents the fraction of particles with $\psi_{4}>0.55$ and $f_{6}$ represents the fraction of particles with $\psi_{6}>0.70$. {\bf b.} Plots of the non-affine parameter $\chi$(green) and the local shear strain component $\epsilon_{s}$(purple) obtained from the best fit affine deformation tensor ${\mathsf D}$. The line representing $\chi$ also shows the $~10$ second time windows into which the whole transformation time has been {\color{black} divided}. Note that all the the parameters $f_{4},f_{6},\chi,\epsilon_{s}$ shows a jump at $time=50 s$. $\chi$ shows a second smaller jump at $time=120 s$.}
\end{center}
\end{figure} 

\begin{figure*}[!]
\begin{center}
\includegraphics[width=0.98\textwidth]{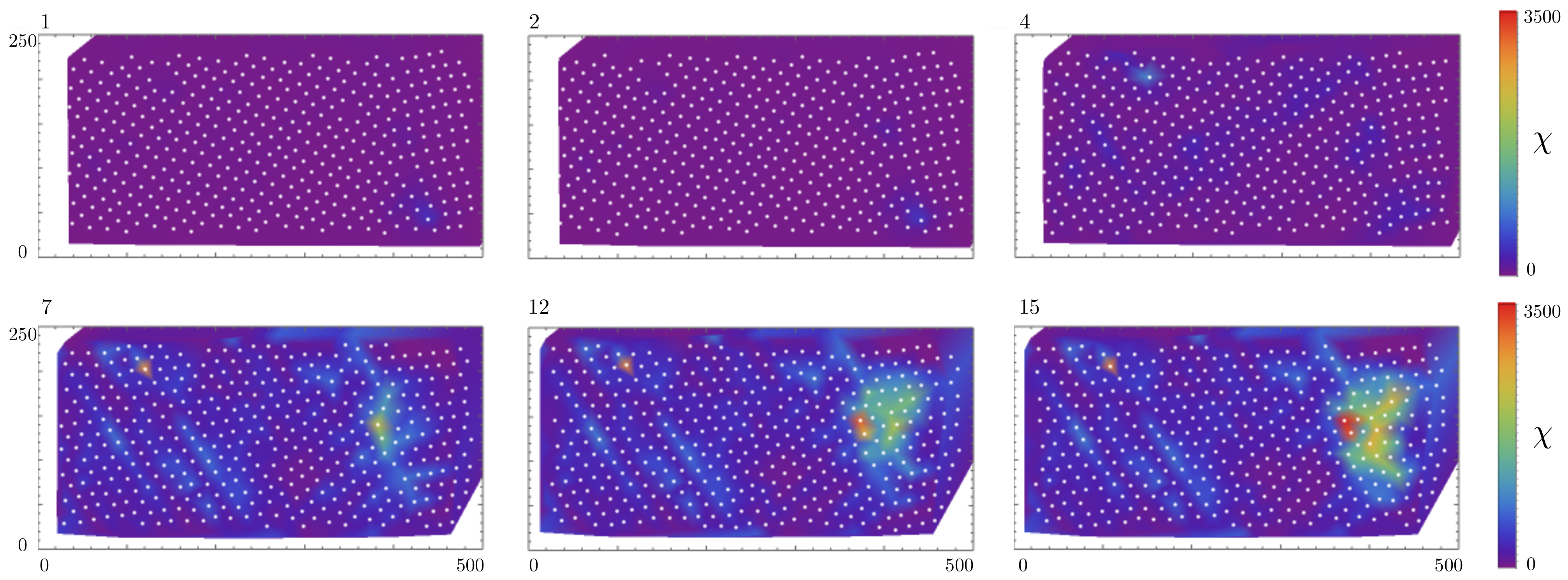}
\caption{\label{chi_map}Averaged configurations at time windows $1, 2, 4, 7, 12, 15$ respectively shown as white symbols superimposed on the averaged non-affine parameter $\chi$ represented as a spatially varying field coded with colors ranging from purple to orange with increasing values of $\chi$. Each configuration was obtained by averaging over 140 snapshots collected over time windows of 10 seconds. Note that the fraction of particles with relatively larger $\chi$ starts to increase at 4 with further local rearrangement, reflected in large local non-affinities, at 12, finally reaching a steady state for the meta stable BCO crystal in 15.}
\end{center}
\end{figure*} 
 Here $N$ is the number of nearest neighbors of a lattice point, $s$ can be either 4 or 6 and $\theta_{j}$ is the angle between the line connecting a nearest neighbor to a lattice point and a reference axis.
The orientation order parameter $\psi_{4}$ and $\psi_{6}$ represent {\color{black} 4-fold and 6-fold coordination symmetries} respectively. In { Ref.~[\citen{Priti}]}, a sharp decrease in the population fraction~($f_{4}$) of 4-fold particles and increase in the 6-fold particles~($f_{6}$) is observed at the transition.  {\color{black} The growth (Avrami) exponent $\alpha$ obtained from the time evolution curve of $f_{6}$ was found to be $8.5$. This is not consistent with the nucleation and 3D grain growth model which dictates an exponent of $\alpha=4$. The coordinated non-diffusive motion of the colloidal particles observed in the CLSM videos also clearly indicates a Martensitic transformation.} We show the evolution of $f_{4}$ and $f_{6}$ with time (calculated from the data in~[\citen{Priti}]) in ~Fig~\ref{Op_time}{\bf a}, as a standard for comparing our new observations.

In ~Fig~\ref{Op_time}{\bf b}, we present $\chi$ (green), in units of $d^2$, as an effective local order parameter for measuring different aspects of the particle displacements and orientations during the course of the stable BCT to {\color{black} metastable} BCO structural transformation. The non-affine parameter, $\chi$ and the shear strain component, $\epsilon_{s}=(\epsilon_{xy}+\epsilon_{yx})/2$ (Fig~\ref{Op_time}{\bf b}, purple curve), of the affine deformation tensor reveals certain distinctive aspects of the particle rearrangements during the transition. 
$\chi$ and $\epsilon_{s}$, both shown in Fig~\ref{Op_time}{\bf b} as a function of time, are calculated with respect to the BCT structure at time $(t=0)$ from the same data set as Fig.\ref{Op_time}{\bf a}.  Similar to $f_{4}$ and $f_{6}$, they show very prominent change at around time ($t=50$ s) indicating the occurrence of the maximum particle rearrangements. $\chi$ also shows a pronounced change at time $t=120$ s. We show below and in the configuration plots in Fig~\ref{chi_map} that this is consistent with the particles still undergoing small local rearrangements which are seen as a very weak blip in the orientation order parameter and the corresponding population fraction $f_4$ (see Fig.\ref{Op_time}{\bf a}, purple curve) but can be observed very clearly using the local non-affinities.

We divide the total transformation time into 15 time windows with each $~ 10$ second window represented by 140 snapshots of configurations. These windows are marked in Fig.\ref{Op_time}{\bf b} with numbers $1-15$. The time windows (1-3) correspond to the BCT structure before transition. The maximum structural rearrangements occur in windows 4 and 5 leading to a still evolving BCO structure~(windows 6-11) with a final spurt of rearrangement in window 12. Fig.~\ref{chi_map} shows configurations at some of the time windows.  Each configuration is obtained by averaging over 140 snapshots collected over time windows of 10 seconds. Here averaged particle positions, represented as white symbols, are superimposed on a spatially varying  $\chi$ {\em field} obtained by interpolating $\chi$ values available at particle positions. The fraction of particles with relatively larger $\chi$ starts to increase at time window 4 getting steady through 5-11 with further local rearrangement, reflected in large local non-affinities, at 12. The configurations 13 to 15 look almost identical owing to much larger relaxation time scales of the meta stable BCO crystal.

We have shown here that the non-affine parameter serves as a highly sensitive local parameter useful in precise detection of rearrangements occurring at an atomic (or single-particle colloid) length scale. In the next section we study these rearrangements and their consequences as the transformation proceeds.





\subsection{Non-affine rearrangements during the transformation}
\label{sec2B}

\begin{figure}[h!]
\begin{center}
\includegraphics[width=0.38\textwidth]{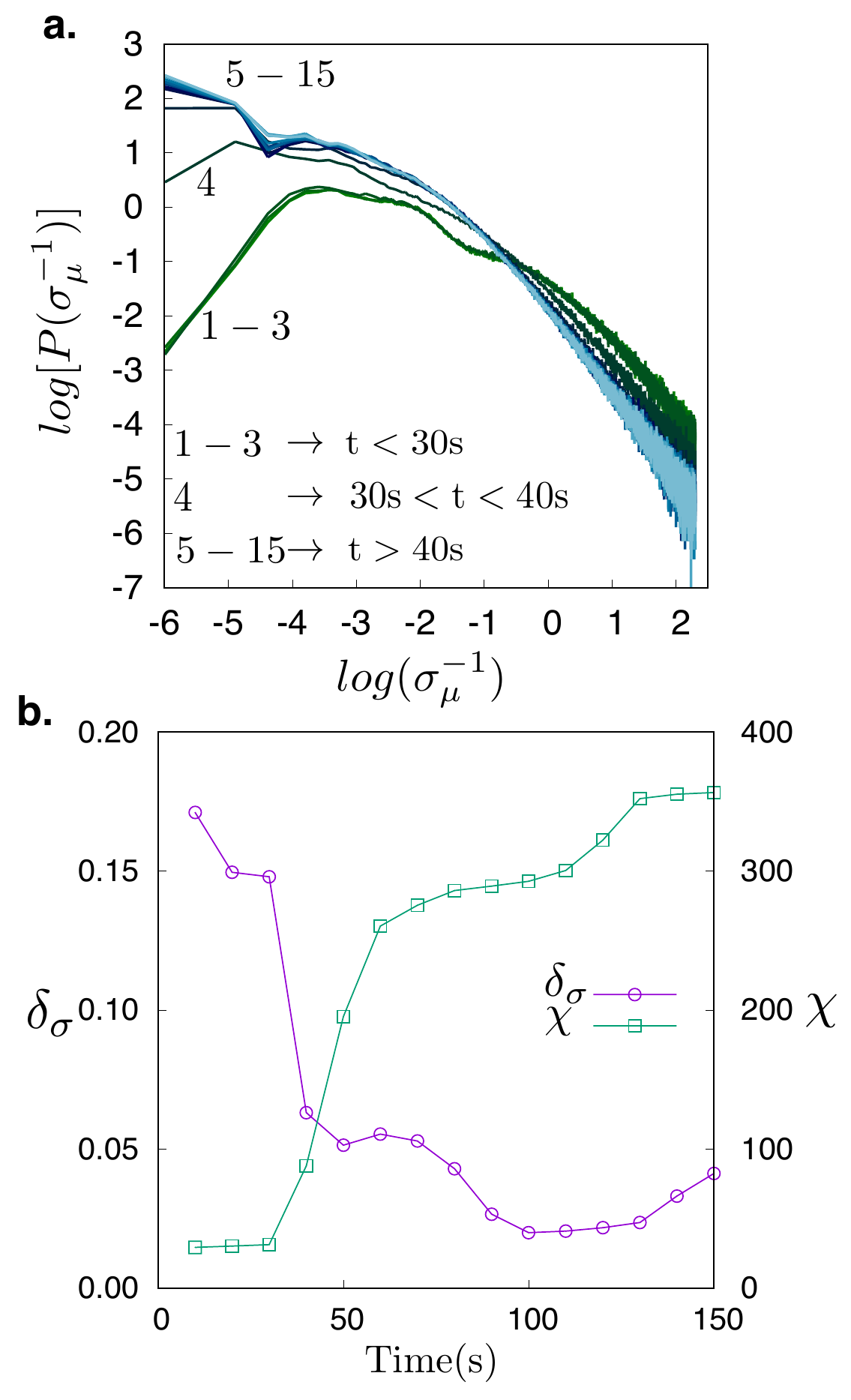}
\caption{
{\bf a} \label{log_pev} $\log[P(\sigma_{\mu}^{-1})]$ plotted as a function of $\log(\sigma_{\mu}^{-1})$. Note the increase in the probability of smaller $\sigma_{\mu}^{-1}$ in line $4$, representing the lattice during the transition as compared to $1-3$ before transition. {\bf b} The gap, $\delta_\sigma$, between the inverse of the eigenvalues of the two most probable non-affine modes plotted as a function of time. Each point is obtained by averaging over 140 snapshots collected over a time window of 10 seconds. Note that its value at the transition is much less than its initial value in the crystalline structure indicating that all the non-affine modes become important during the transition. The evolution of $\chi$ is also plotted for reference.}
\end{center}
\end{figure} 

As stated earlier, the non-affine eigenvalues $\sigma_{\mu}$ can be interpreted as the susceptibilities associated with specific non-affine modes: smaller $\sigma_{\mu}^{-1}$ implies a softer mode.
The probability distributions, $P(\sigma_{\mu}^{-1})$~(Fig.~\ref{log_pev}{\bf a}) of the inverse of these non-affine eigenvalues ($\sigma_{\mu}^{-1}$) have been calculated from configurations at various time windows~(see Section~\ref{sec2A} and Fig.~\ref{Op_time}) at the various stages of the transformation.  {\color{black}For an ideal crystal at the thermodynamic limit, this distribution would have delta peaks centered around the $\sigma_{\mu}^{-1}$s shown in Fig.~\ref{Ev_Harm}{\bf b.} Finite system size, introduction of dislocations, inaccuracies arising from taking 2D cross sections of a 3D crystal and limited statistics of the experimental data leads to broadening and overlapping of the individual  peaks. The distributions corresponding to the parent BCT structure in time windows 1,2,3 in Fig.~\ref{log_pev}{\bf a.} shows two broad peaks. The inverse of the softest non-affine modes like $\sigma_{1}^{-1}$ contribute to the first peak while the rest of the non-affine modes contribute to the second broad peak and the tail of the distribution.}
The $P(\sigma_{\mu}^{-1})$ for the fourth and the subsequent time windows in Fig.~\ref{log_pev}{\bf a} shows an increased probability of softer (smaller $\sigma_{\mu}^{-1}$) non-affine modes compared to the $P(\sigma_{\mu}^{-1})$ for the BCT structure. This reflects the greater ease (and hence large numbers) of non-affine rearrangements occurring at this stage of the transformation~(Fig.\ref{chi_map}).

We also estimated the gap ($\delta_{\sigma}=\sigma_{2}^{-1}-\sigma_{1}^{-1}$) between the inverse of the eigenvalues of the two most probable non-affine modes (Fig.~\ref{log_pev}{\bf b}) as a function of the transformation time. {\color{black}The gap $\delta_{\sigma}$, for the model harmonic square network is $0.96$ (Fig.~\ref{Ev_Harm}{\bf b.}) and this sets an upper bound for this quantity. The value of $\delta_{\sigma}$ obtained from the 2D confocal images of the quiescent BCT crystal is $\approx 0.175$. The smaller value of the $\delta_{\sigma}$ from the experimental data can be attributed to the same factors that lead to the broadening of the $P(\sigma_{\mu}^{-1}$). As the crystal structure starts to undergo rearrangements, thus moving away from the parent BCT structure, during the Martensitic transition, the $\delta_\sigma$ shows a marked decrease (Fig.~\ref{log_pev}{\bf b}). The gap $\delta_{\sigma}$ separates {\color{black} preferred non-affine modes from the rest of the spectrum.  As mentioned before,~\cite{SG2} these are the defect precursors, which in an ideal crystal at finite temperature are the only significant non-affine fluctuations.} So a markedly lower $\delta_{\sigma}$ from the quiescent BCT value of close to 0.2 to a value close to 0.03, indicates that during the BCT to BCO transformation, there is an almost equal participation of all the non-affine modes. {\color{black} Consequently, a crystal undergoing structural transformation cannot be unambiguously represented as an elastic continuum with a finite density of defects. On the other hand, characterizing the displacement spectrum in terms of specific non-affine modes makes it possible to quantify the precise difference in displacements in a crystalline phase and the transforming structures.}  
%

\subsection{Identification of different transformation paths leading to similar crystalline structure}
\label{sec2C}
\begin{figure*}[!]
\begin{center}
\includegraphics[width=0.98\textwidth]{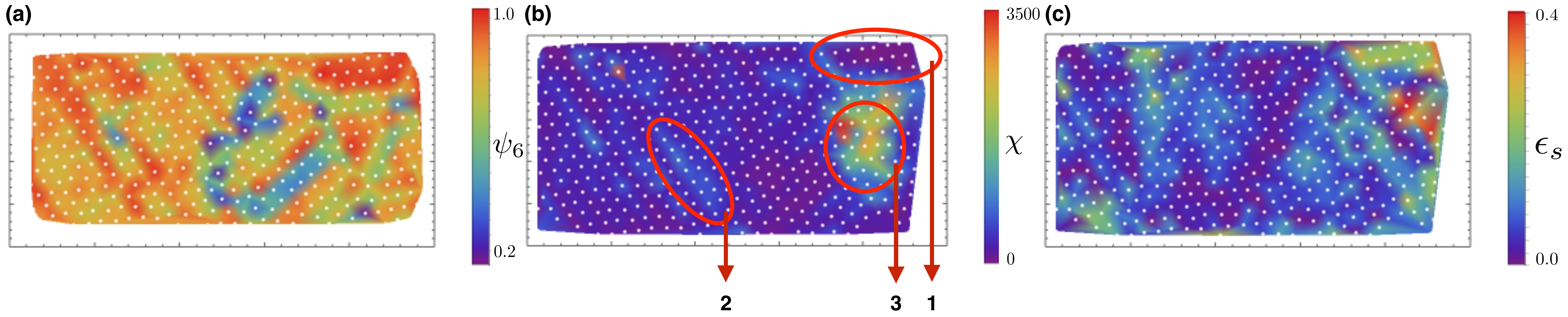}
\caption{ \label{Diff_history}{\color{black} Averaged configurations together with superimposed fields shown as colour maps as in Fig.~\ref{chi_map} {\bf a.}Map of $\psi_{6}$ together with particle configurations} at $time=150$ seconds. A predominant 6-fold(BCO) symmetry is observed. {\bf b.}$\chi$ map of the same configuration, pointing out regions 1, 2, 3 with varying amounts of non-affine displacements. {\bf c.}$\epsilon_{s}$ map of the same configuration with varying amounts of shear strain. Remarkably, region (1) with almost zero non-affine displacements has large affine shear strain, region (2) has small shear strains as well as small non-affine displacements and region (3)has undergone much larger non-affine displacements with comparably small affine shear component. Note that the three regions has similar values of $\psi_{6}$.}
\end{center}
\end{figure*}

{\color{black} In this subsection we present results which show unambiguously that different regions of the {\em same} product BCO phase with identical final crystalline order can be, nevertheless, formed through {\em either} almost purely affine shear transformations or relatively large non-affine rearrangements.} The local shear strains and the non-affinities are estimated with respect to the parent BCT structure.

The bond orientational order parameters $\psi_{4}$ and $\psi_{6}$ are used in identifying the co-ordination symmetries of particles. Particles with high  $\psi_{4}$ and $\psi_{6}$ (in~[\citen{Priti}], the thresholds were $\psi_{4}>0.55$ and $\psi_{6}>0.7$) are considered to have 4 (BCT) and 6 (BCO) coordination numbers respectively. Local $\psi_{6}, \chi$ and $\epsilon_{s}$ are calculated for particles in the configuration obtained at $time=150$ seconds. Fig.~\ref{Diff_history}(a), shows a colour map of $\psi_{6}$. 
Although there are slight spatial variations in the values of $\psi_{6}$, the overall coordination symmetry of the lattice is represented by the BCO structure. The local $\chi$ map (Fig.~\ref{Diff_history}(b)) of the same configuration shows three regions with very different values of local non-affinities - region (1) has undergone almost no non-affine displacements while region (3) has undergone very large non-affine displacements. However, the affine shear strain ($\epsilon_{s}$) map in Fig.~\ref{Diff_history} (c) indicates that region (1) has undergone much larger affine transformation compared to region (2) and (3).  This clearly depicts that regions with identical crystalline order can be attained through purely affine shear transformation as in case of region (1) or predominantly non-affine particle displacements as in region(3). The local non-affine parameter $\chi$ therefore depicts not only the current structure of a neighbourhood but also its history. We believe that this analysis would be extremely useful for studying structural transitions where such detailed information is needed.

\section{Discussion and conclusions}
\label{sec3} 

{\color{black} In this work, we find that the quantification of affine and non-affine deformations is a natural and sensitive tool for studying solid-solid phase transformations in colloids. They are natural, because, the particle rearrangements are the primary events in the phase transformation process. They are seen in this work to be also highly sensitive indicators}; for example,
the hot spot in Fig.~\ref{chi_map}~(time window 12) correlates with the jump in $\chi$ in Fig.~\ref{Op_time}{\bf b} for $t > 100$ seconds. This jump is, on the other hand, nearly undetectable using bond order parameters fractions or maps (Fig.~\ref{Op_time}{\bf a} or Fig.~\ref{Diff_history}{\bf a} respectively). We thus suggest that {\it any} colloidal phase transformation may benefit from similar analysis techniques.

In Ref.~[\citen{SG2}], it has been demonstrated that the non-affine displacement modes with the largest susceptibilities (hence the softest and most probable) are those which tend to create defects precursors viz. dislocation dipoles. 
In the crystalline solid these defect precursor modes are energetically well separated from the rest of the thermal non-affine displacement modes by a large {\em gap} in the non-affine excitation spectrum. {\color{black}The colloidal Martensitic transformation from the body-centred tetragonal (BCT) crystal to a metastable body-centred orthorhombic (BCO) crystal is accompanied by two main effects:
\begin{enumerate}
\item  an increase in the density of softer non-affine modes(Fig.~\ref{log_pev}{\bf a.}). 
\item a marked decrease in the gap in the non-affine excitation spectrum (Fig.\ref{log_pev}{\bf b.}).
\end{enumerate}}
\noindent Effect (1) implies a greater ease of particle rearrangements. Effect (2), on the other hand, reflects an almost equal participation of all the non-affine modes during the transformation phase of the solid. 
The latter is a distinctive feature of the displacement spectrum of the transforming phase compared to that of a finite temperature crystal structure which is predominated by only the most probable or {\it defect-precursor-}like non-affine modes. 

The non-affine parameter also codes the history of the transformation. {\color{black}Considering the nature of a Martensitic transformation, one would expect that the transformation within a pure crystal grain will be predominated by simple "affine" displacements with small changes in particle neighbourhoods. However, a different local elastic environment at the defects or grain boundaries in the reference parent crystal might make them more prone to rearrangements which cannot be represented simply in terms of affine deformations.  As $\chi$ is sensitive to the precise nature of the displacements in particle neighbourhoods,  $\chi$ will tend be larger at regions of defects and grain boundaries. This is nicely depicted by the large non-affine displacements in the region 3 in Fig.~\ref{Diff_history}{\bf b}, which coincides with the position of the grain boundary seen in the parent crystal in Fig.~\ref{chi_map}(1,2,4).} Careful analysis of the non-affine displacements in the meta-stable BCO product crystal makes it possible to identify distinct regions which have identical crystal symmetries. One can associate these regions with very distinct values of local shear strains and non-affinities. There are regions with large values of affine shear strains and very small non-affine displacements as well as regions, with small shear strains, while it has undergone large non-affine displacements. Thus this approach makes it possible to identify regions formed through different transformation paths even though their final structures are identical. We observe that the final product may be formed by both elastic strain (shear), as well as, particle rearrangements in different portions of the sample - making a ``strain-only" description of this transition essentially incomplete~\cite{Barsch,Lookman,Lookman_Shenoy1,Gooding}.

{\color{black}Theories dealing with smooth displacement fields fail to give a complete description of structural transitions accompanied by plastic deformations. As we have shown, all the non-affine displacement modes play an important role. So introducing dislocations alone also provides a partial picture for the plastic part of the displacements. It is difficult to separate out transformations which are purely affine and purely diffusive because both may occur in the same system and over roughly the same time scales. This necessitates a  much generalized theory capable of addressing the various facets of structural transition. Our present work provides some useful insights into the microscopic mechanism of such transitions. We intend to extend our ideas for similar colloidal systems where carefully controlled experiments are possible.}

\section{Acknowledgements}
Illuminating discussions and suggestions by J\"urgen Horbach and Subodh Shenoy are gratefully acknowledged. SG and SS acknowledge funding from the FP7-PEOPLE-2013-IRSES grant no: 612707, DIONICOS. PSM acknowledges partial support from SERB, Department of Science and Technology (Ref:   SB/S3/CE/042/2015), India. PS acknowledges financial support from the Swedish Research Council (Project 2011-4338) and the European Research Council (ERC-339678-COMPASS). A.Y. acknowledges funding support from the Natural Sciences and Engineering Research Council of Canada. A.Y. also thanks members of TCIS for their generous hospitality during the course of the studies.

\bibliography{SuSe-SG_colloid}


\end{document}